\documentclass[twocolumn,flushbottom,showpacs,preprintnumbers,superscriptaddress,amsmath,amssymb]{revtex4}
\usepackage{graphicx}
\usepackage{dcolumn}
\usepackage{bm}
\usepackage{amssymb}
\usepackage{amssymb}
\def\>{\rangle}

\begin{document}

\title{Non-Markovian quantum input-output networks}

\author{Jing Zhang}\email{jing-zhang@mail.tsinghua.edu.cn}
\affiliation{Department of Automation, Tsinghua University,
Beijing 100084, P. R. China} \affiliation{Center for Quantum
Information Science and Technology, TNList, Beijing 100084, P. R.
China} \affiliation{Advanced Science Institute, RIKEN, Wako-shi,
Saitama, 351-0198, Japan}
\author{Yu-xi Liu}
\affiliation{Institute of Microelectronics, Tsinghua University,
Beijing 100084, P. R. China} \affiliation{Center for Quantum
Information Science and Technology, TNList, Beijing 100084, P. R.
China} \affiliation{Advanced Science Institute, RIKEN, Wako-shi,
Saitama, 351-0198, Japan}
\author{Re-Bing Wu}
\affiliation{Department of Automation, Tsinghua University,
Beijing 100084, P. R. China} \affiliation{Center for Quantum
Information Science and Technology, TNList, Beijing 100084, P. R.
China} \affiliation{Advanced Science Institute, RIKEN, Wako-shi,
Saitama, 351-0198, Japan}
\author{Kurt Jacobs} \affiliation{Department of Physics, University of Massachusetts at Boston, Boston, MA 02125, USA}
\affiliation{Advanced Science Institute, RIKEN, Wako-shi, Saitama,
351-0198, Japan}
\author{Franco Nori}
\affiliation{Advanced Science Institute, RIKEN, Wako-shi, Saitama,
351-0198, Japan} \affiliation{ Physics Department, The University
of Michigan, Ann Arbor, Michigan 48109-1040, USA}

\date{\today}

\begin{abstract}
Quantum input-output response analysis is a useful method for
modeling the dynamics of complex quantum networks, such as those
for communication or quantum control via cascade connections.
Non-Markovian effects have not yet been studied in such networks.
Here we extend the Markovian input-output network formalism
developed in optical systems to non-Markovian cascaded networks
which can be used, e.g., to analyze the input-output response of
mesoscopic quantum networks. We use this formalism to explore the
behavior of superconducting qubit networks, where we examine the
effect of finite cavity bandwidths. We also discuss its
application to open- and closed-loop control networks, and show
how these networks create effective Hamiltonians for the
controlled system.
\end{abstract}


\maketitle

\section{Introduction}\label{s1}
There has been tremendous progress in the last few years in
experimental efforts to realize quantum networks~\cite{Kimble} in
various mesoscopic systems. These systems include photonic
crystals~\cite{Obrien}, ion traps~\cite{Ions}, and superconducting
circuits~\cite{SuperC}, which also advancesrelated fields such as
quantum simulation (for recent reviews, see, e.g.,
Ref.~\cite{Quantum_Simulator}). The input-output formalism of
Gardiner and Collet~\cite{Gardiner,Carmichael} is a useful tool
for analyzing such networks. In fact, using the input-output
response to analyze or even modify the dynamics is a standard
method in engineering called system synthesis. Up until now,
system synthesis for quantum networks has only been studied for
Markovian systems. Quantum input-output theory itself has also
mainly been limited to the Markovian regime, although it was
developed for quantum systems about twenty years
ago~\cite{Gardiner}. It was extended to the non-Markovian systems
only quite recently~\cite{LDiosi}.

In the existing literature~\cite{Gardiner,Carmichael}, quantum
input-output theory is mainly applied to optical systems, in which
the coupling between the system and its environment is weak and
the correlation time (the ``memory'' of the environment) is small
compared with the characteristic time-scale of the system
dynamics. Under the Markovian assumption, the quantum input-output
formalism~\cite{Gardiner} was extended to cascaded
systems~\cite{Carmichael}, and has been used to study quantum
feedforward and feedback networks~\cite{Gough,Nurdin,GFZhang}.
Markovian quantum input-output networks can be described using two
alternative formulations: the Hudson-Parthasarity formalism in the
Schr\"{o}dinger picture~\cite{Hudson}; and the quantum transfer
function formalism in the Heisenberg
picture~\cite{MYanagisawa1,Gough2}. The general algebraic
structure of such systems has been well studied in the language of
quantum Wiener and Poisson processes and quantum Ito
rules~\cite{Hudson}.

Although the Markovian assumption is reasonable when considering
optical network components, environments in mesoscopic solid-state
systems can have correlations on much longer
timescales~\cite{Nazir,CuiWei,ShiBei,Wiseman3}. Examples of this
are the nuclear spin bath that couples to electron spins in
quantum dots~\cite{Marcus}, and the $1/f$ noise that affects
Josephson-Junction qubits~\cite{McDermott}. It has also been
suggested that the damping and decoherence of nano-mechanical
resonators are due to coupling to a small number of two-level
systems~\cite{Miles}, which can be expected to induce significant
non-Markovian dynamics. In addition, any classical noise with a
sufficiently narrow band generates non-Markovian evolution.
There has been increasing interest in recent years in non-Markovian open quantum
systems, and a number of analytical approaches have been devised to describe
them. These include the projection-operator partitioning technique~\cite{Breuer}, the
non-Markovian quantum trajectory approach~\cite{WTStrunz}, and very recently
a non-Markovian input-output formalism~\cite{LDiosi}.

In this paper we extend the non-Markovian input-output theory to
cascaded quantum networks, providing a recipe for obtaining
non-Markovian input-output equations for the description of any
such network. Naturally, this formalism reduces to the standard
input-output network formalism in the Markovian limit. Although
the non-Markovian input-output relation has been derived in
Ref.~\cite{LDiosi}, a quantum measurement is imposed on the output
field in Ref.~\cite{LDiosi} and the system dynamics is described
by the quantum state diffusion equation. Such a formalism cannot
be extended to describe a non-Markovian cascade system, because
quantum coherence in the output field is deteriorated by the
measurement. In our formalism, without introducing measurements,
the system dynamics is described by a non-Markovian quantum
stochastic differential equation and a perturbative master
equation which can be naturally extended to a non-Markovian
network.

In developing our non-Markovian formalism, we will keep the
weak-noise (weak-coupling) approximation, as this is the
appropriate regime for implementing quantum technologies (for
recent reviews, see, e.g., Ref.~\cite{Buluta}), such as
information processing and metrology. Here, we go beyond the
Markovian approximation by allowing the coupling to the bath to
have an arbitrary frequency dependence. This allows one to
describe noise with any frequency profile, and should provide a
good model for a wide range of non-Markovian environments.
Interestingly, the resulting non-Markovian network formalism is
exact for all couplings. However, in order to perform calculations
for nonlinear systems, one must transform the Heisenberg equations
of the input-output formalism to non-Markovian master equations,
and this requires further approximations. Here we do this at the
simplest level of approximation, by deriving the corresponding
master equation to second-order in perturbation theory, using the
standard Born approximation~\cite{Breuer}. Nevertheless, more
sophisticated techniques exist for obtaining non-Markovian master
equations, and it would be an interesting avenue for future work
to examine how these can be used to obtain master equations for
non-Markovian cascaded networks. We note that for linear systems
the Heisenberg equations of the input-output formalism can be used
to obtain exact results. This is especially useful in some
cases~\cite{Vitali}, when the second-order perturbative master
equation fails to behave correctly~\cite{Jacobs4}. We expect the
network formalism we develop to be useful in describing a range of
mesoscopic systems, such as coupled-cavity arrays in photonic
crystals~\cite{ZhangWM,CPSun}, and nonlinear resonator and qubit
networks in solid-state
circuits~\cite{SC,Jacobs2,Mabuchi2,JZhang}. The formalism can also
be applied to quantum feedback control
networks~\cite{SLloyd,Mabuchi,Wiseman1,Doherty1,Belavkin,Jacobs1,ZHYan}
in solid-state
systems~\cite{YYamamoto,AYacoby,ANKorotkov,TBrandes,Jacobs3,GJMilburn,XQLi,JZhang2}.


This paper is organized as follows: in Sec.~\ref{s2}, we briefly
review the Markovian input-output formalism so that this can be
easily compared to the non-Markovian case. In Sec.~\ref{s3}, we
use an alternative method to derive the non-Markovian input-output
relations in Ref.~\cite{LDiosi}, and obtain the dynamical equation
for the system such that it can be easily used for networks. We
derive these here as a natural extension of the original
Collett-Gardiner quantum input-output theory. In Sec.~\ref{s4}, we
derive the input-output relations for more complex non-Markovian
quantum cascade networks. We then apply these general results to
two examples: a simple non-Markovian oscillator, and a network of
two superconducting charge qubits interacting with Lorentz noises.
We also apply the formalism to an open-loop and a closed-loop
control networks, showing how these networks create effective
Hamiltonians for the controlled system. The conclusions and
prospects for future work are discussed in Sec.~\ref{s5}.

\section{Brief review of input-output theory of Markovian systems}\label{s2}

Here we summarize the standard Gardiner-Collet input-output
formalism~\cite{Gardiner}. The basic model is a quantum system
interacting with a bath, where the bath consists of the modes of
an electromagnetic field, or equivalently a continuum of harmonic
oscillators. The Hamiltonian for the system and bath is
\begin{eqnarray}\label{General Hamiltonian}
H&=&H_S+H_B+H_{\rm int},
\end{eqnarray}
with
\begin{eqnarray}
H_B&=&\int_{-\infty}^{+\infty}\!\!\! \omega\;b^{\dagger}\!\left(
\omega \right) b\!\left( \omega \right) d\omega,\nonumber\\
H_{\rm int}&=& i \int_{-\infty}^{+\infty}\!\!\! \left[ \kappa
\left( \omega \right) b^{\dagger}\left( \omega \right) L - {\rm
h.c.} \right] d \omega,
\end{eqnarray}
where $b^{\dagger}\left(\omega\right)$ and $b \left( \omega
\right)$ are the creation and annihilation operators of the bath
mode with frequency $\omega$, which
satisfy
\begin{equation}\label{Commutation relation of continuous
variable}
\left[b\left(\omega\right),b^{\dagger}\left(\tilde{\omega}\right)\right]=\delta\left(\omega-\tilde{\omega}\right).
\end{equation}
In the above, $H_S$ is the free Hamiltonian of the system. The
bath mode with frequency $\omega$ interacts with the system via
the system operator $L$ and the coupling strength $\kappa \left(
\omega \right)$. Hereafter we set $\hbar=1$. The total Hamiltonian
$H$ can be reexpressed in the interaction picture as
\begin{eqnarray}\label{Effective Hamiltonian of the total system}
H_{\rm eff} & = & \exp{ \left( i H_B t \right) } \left( H_S +
H_{\rm int} \right) \exp{ \left( -i H_B
t \right) }\nonumber\\
& = & H_S+i\int_{-\infty}^{+\infty}\!\!\! \left[ \kappa \left(
\omega \right) e^{i \omega t} b^{\dagger} \left( \omega \right) L
-{\rm h.c.}\right] d \omega.
\end{eqnarray}
If the coupling strength is constant for all frequencies, so that
\begin{equation}\label{Markovian assumption}
\kappa \left( \omega \right) = \sqrt{ \frac{\gamma}{2\pi} },
\end{equation}
then the dynamics of the system will become Markovian. This is the
Markovian approximation. The Hamiltonian $H_{\rm eff}$ is now given by
\begin{equation}\label{Effective Hamiltonian under the Markovian assumption}
H_{\rm eff}=H_S+i \sqrt{\gamma} \left[ b_{\rm in}^{\dagger} \left(
t \right) L - L^{\dagger} b_{\rm in} \left( t \right) \right],
\end{equation}
where
\begin{equation}\label{Input field}
b_{\rm in}\left( t
\right)=\frac{1}{\sqrt{2\pi}}\int_{-\infty}^{+\infty}\!\!\! e^{-i
\omega t} b \left( \omega \right) d\omega
\end{equation}
is the Fourier transform of the bath modes, and is the
time-varying input field that is fed into the system~\cite{KJPhD}
(see Fig.~\ref{Fig of the Markovian input-output system}). In the
Heisenberg picture, the system operator $ X \left( t \right) $
satisfies the following quantum stochastic differential equation
(QSDE)
\begin{eqnarray}\label{Quantum stochastic differential equation}
\dot{X}&=&-i\left[X,H_S\right] + \frac{\gamma}{2} \left\{
L^{\dagger} \left[ X,
L \right] + \left[ L^{\dagger}, X \right] L \right\} \nonumber \\
&& + \sqrt{\gamma} \left\{ b_{\rm in} \left[ L^{\dagger}, X
\right] + \left[ X, L \right] b_{\rm in}^{\dagger} \right\}.
\end{eqnarray}
If the input field is in a $b_{\rm in} \left( t \right)$ vacuum state, and we trace it out,
we can re-express the system dynamics in the Schr\"{o}dinger picture as the following
master equation
\begin{equation}\label{Master equation of the Markovian input-output system}
\dot{\rho}=-i \left[ H_S, \rho \right] + \gamma \left( L \rho
L^{\dagger} - \frac{1}{2} L^{\dagger}L \rho -\frac{1}{2} \rho
L^{\dagger} L \right).
\end{equation}
Finally, if $b_{\rm out} \left( t \right)$ is the field after it
has interacted with the system, and is now out of it, then one has
the relation
\begin{equation}\label{Output equation under the Markovian approximation}
b_{\rm out} \left( t \right) = b_{\rm in}(t) + \sqrt{\gamma} L(t).
\end{equation}
This is the Markovian input-output relation.

\begin{figure}[t]
\centerline{\includegraphics[width=6.45 cm]{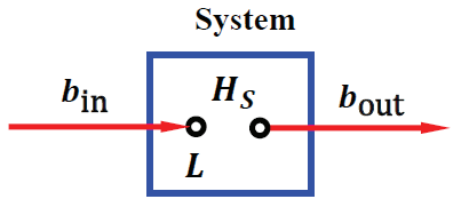}}
\caption{(Color online) Schematic diagram of the Markovian
input-output system.}\label{Fig of the Markovian input-output
system}
\end{figure}

\section{Input-output theory of non-Markovian systems}\label{s3}

\subsection{General theory}\label{s31}

To derive the input-output relation for a general non-Markovian
quantum system (see Fig.~\ref{Fig of the non-Markovian
input-output system}), we rewrite the Hamiltonian $H_{\rm
eff}$ in Eq.~(\ref{Effective Hamiltonian of the total system}) as
\begin{eqnarray}\label{Effective Hamiltonian for non-Markovian system}
H_{\rm eff} & = & H_S + i \left\{ \left[
\int_{-\infty}^{+\infty}\!\!\! \kappa\!\left( \tau - t \right)
b_{\rm in}^{\dagger} \left(
\tau \right) d \tau \right] L - {\rm h.c.} \right\} \nonumber \\
& = & H_S + i \left[ \tilde{b}_{\rm in}^{\dagger} \left( t \right)
L - L^{\dagger} \tilde{b}_{\rm in} \left( t \right) \right],
\end{eqnarray}
where
\begin{equation}\label{Fourier transform of the coupling strength}
\kappa \left( t \right) = \frac{1}{\sqrt{2 \pi}} \int_{-
\infty}^{+ \infty}\!\!\! \exp{ \left( -i\omega t \right) } \kappa
\left( \omega \right) d \omega
\end{equation}
is the Fourier transform of the coupling strength $\kappa \left(
\omega \right)$. The input field that interacts directly with the system
is now
\begin{equation}\label{Equivalent non-Markovian input field}
\tilde{b}_{\rm in} \left( t \right) = \int_{ - \infty }^{ + \infty
}\!\!\! \kappa\!\left( t - \tau \right) b_{\rm in}\!\left( \tau
\right) d \tau
\end{equation}
and satisfies the new commutation relation
\begin{equation}\label{Commutation relation of the non-Markovian input field}
\left[ \tilde{b}_{\rm in} \left( t \right), \tilde{b}_{\rm
in}^{\dagger} \left( \tilde{t} \right) \right] = \gamma \left( t -
\tilde{t} \right),
\end{equation}
where
\begin{equation}\label{Non-Markovian correlation rate}
\gamma \left( t - \tilde{t} \right) = \int_{ - \infty }^{ + \infty
}\!\!\! \kappa^* \left( t - \tau \right) \kappa \left( \tilde{t} -
\tau \right) d \tau.
\end{equation}
\begin{figure}[t]
\centerline{\includegraphics[width=6.45 cm]{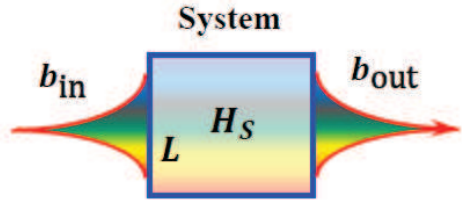}}
\caption{(Color online) Schematic diagram of the non-Markovian
input-output system. The input field is dispersed when it
interacts with the system, and the modes with different
frequencies in the input field are coupled to the system with
different coupling strengths.}\label{Fig of the non-Markovian
input-output system}
\end{figure}

We can now proceed to derive the Heisenberg stochastic
differential equations for the evolution of the system:
\begin{eqnarray}\label{Quantum stochastic differential equation for non-Markovian system}
\dot{X}&=&-i\left[X,H_S\right]+\int_0^t\!\!\! \left\{
\gamma\left(t-\tau\right) L^{\dagger} \left( \tau \right) \left[ X
\left( t \right), L
\left( t \right) \right] \right. \nonumber \\
&&\left. + \gamma^* \left( t - \tau \right) \left[ L^{\dagger}
\left( t \right), X \left( t \right)
\right] L \left( \tau \right) \right\} d \tau \nonumber \\
&& + \left\{ \tilde{b}_{\rm in} \left( t \right) \left[
L^{\dagger}, X \right] + \left[ X, L \right] \tilde{b}_{\rm
in}^{\dagger} \left( t \right) \right\}.
\end{eqnarray}
The non-Markovian input-output relation becomes
\begin{equation}\label{Output equation of the non-Markovian system}
b_{\rm out}\left( t \right) = b_{\rm in} \left( t \right) +
\int_0^t\!\!\! \kappa \left( t - \tau \right) L \left( \tau
\right) d \tau.
\end{equation}
Note so far that no further assumptions have been made, based on
the Hamiltonian in Eq.(\ref{Effective Hamiltonian for
non-Markovian system}). Thus our non-Markovian network formalism
will give a more precise description of non-Markovian systems with
environment's coupling that have an arbitrary
frequency-dependence. However, as noted in the introduction, for
non-linear networks we must often resort to the Schr\"{o}dinger
picture to perform calculations.

To obtain the second-order perturbative master equation, one averages
over the vacuum input field $b_{\rm in}$, which we will take to be in the vacuum
state, and uses the Born approximation. The perturbative master equation
that corresponds to Eq.~(\ref{Quantum stochastic differential equation for
non-Markovian system}) is
\begin{eqnarray}\label{Non-Markovian master equation}
\dot{\rho} & = & -i \left[ H_S, \rho \right] + \int_0^t\!\!\!
\left\{ \gamma \left( t - \tau \right)  \left[ L \rho \left( \tau
\right), L_{\rm H_S}^{\dagger} \left( \tau - t \right) \right]
\right. \nonumber\\
&&\left. + \gamma^* \left( t - \tau \right) \left[ L_{\rm H_S}
\left( \tau - t \right), \rho \left( \tau \right) L^{\dagger}
\right] \right\} d \tau,
\end{eqnarray}
where
\begin{equation}\label{LHS}
L_{\rm H_S} \left( t \right) = \exp{ \left( i H_S t \right) } L
\exp{ \left( -i H_S t \right) }.
\end{equation}
We give the details of the derivations of Eqs.~(\ref{Quantum
stochastic differential equation for non-Markovian system}) and
(\ref{Non-Markovian master equation}), and output
equation~(\ref{Output equation of the non-Markovian system}) in
Appendix~\ref{as1}.
\\[0.05cm]

\noindent {\bf Remark 1:} The input-output relation, i.e.,
Eq.(\ref{Output equation of the non-Markovian system}), coincides
with Diosi's non-Markovian input-output equation (Eq.~(10) in
Ref.~\cite{LDiosi}). However, the dynamics of the input-output
system in Ref.~\cite{LDiosi} is described by the quantum state
diffusion equation after introducing a heterodyne measurement on
the output field. Such a formalism cannot be extended to derive
the dynamics of more complex non-Markovian networks, because
quantum effects of the input-output system have been reduced by
the quantum measurements. In our formalism, without introducing
measurement, the system dynamics is described by the non-Markovian
quantum stochastic differential equation~(\ref{Quantum stochastic
differential equation for non-Markovian system}) and the
perturbative master equation~(\ref{Non-Markovian master
equation}). Such a formalism can be more naturally extended to
non-Markovian cascade networks.
\\[0.05cm]

\noindent {\bf Remark 2:} In the Markovian limit, where $\kappa
\left( \omega \right) = \sqrt{ \gamma / 2 \pi }$, we have $ \kappa
\left( t \right) = \sqrt{ \gamma } \delta \left( t \right)$, and
$\gamma\left(t-\tilde{t}\right) =  \gamma \delta \left( t -
\tilde{t} \right)$. It can be easily verified that
Eq.~(\ref{Quantum stochastic differential equation for
non-Markovian system}) reduces to the quantum stochastic
differential equation (\ref{Quantum stochastic differential
equation}) and the output equation (\ref{Output equation of the
non-Markovian system}) reduces to Eq.~(\ref{Output equation under
the Markovian approximation}).
\\[0.05cm]

\subsection{Example: Single-mode cavity}\label{s32}

For a single-mode cavity coupled to
an external input field (see Fig.~\ref{Fig of the linear non-Markovian cavity}), the system
Hamiltonian $H_S$ and dissipation operator $L$ are given by
$H_S = \omega_0 a^{\dagger} a$ and $L=a$, where $\omega_0$, $a$
($a^{\dagger}$) are respectively the frequency and the annihilation (creation)
operator of the cavity mode. The mode has an arbitrary non-Markovian
coupling to the field modes outside the cavity.
\begin{figure}[t]
\centerline{\includegraphics[width=6.65 cm]{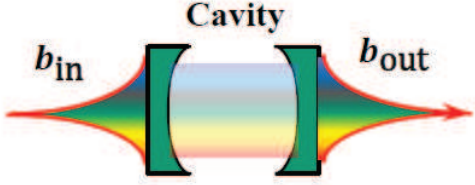}}
\caption{(Color online) Schematic diagram of the linear
non-Markovian cavity.}\label{Fig of the linear non-Markovian
cavity}
\end{figure}
The quantum stochastic differential equation for a cavity operator
$X$ is then
\begin{eqnarray}\label{Quantum stochastic differential equation with single-mode cavity}
\dot{X} & = & - i \left[ X, \omega_0 a^{\dagger} a \right] +
\int_0^t\!\!\! \left\{ \gamma \left( t - \tau \right) a^{\dagger}
\left( \tau \right) \left[ X \left( t \right), a \left( t \right)
\right]
\right. \nonumber \\
& & \left.
+ \gamma^* \left( t - \tau \right) \left[ a^{\dagger} \left( t \right), X \left( t \right) \right] a \left( \tau \right) \right\} d \tau \nonumber \\
& & + \left\{ \tilde{b}_{\rm in} \left[ L^{\dagger}, X \right] +
\left[ X, L \right] \tilde{b}_{\rm in}^{\dagger} \right\}.
\end{eqnarray}
Let us define the normalized position and momentum operators for the
cavity as $q
= \left( a + a^{\dagger} \right) / \sqrt{2}$, $p = \left( -i a + i
a^{\dagger} \right) / \sqrt{2}$. Collecting these into a single vector,
$\vec{x} = \left( q, p \right)^T$, we may write the equations of motion for these operators as
\begin{eqnarray}\label{Dynamical equation of single-mode cavity}
\dot{\vec{x}} & = & A_0 \vec{x} + A_1 \int_0^t\!\!\! \gamma \left(
t -
\tau \right) \vec{x} \left( \tau \right) d \tau \nonumber \\
& & + A_1^* \int_0^t\!\!\! \gamma^* \left( t - \tau \right)
\vec{x} \left( \tau \right) d \tau + B \tilde{b}_{\rm in} + B^*
\tilde{b}_{\rm in}^{\dagger},
\end{eqnarray}
where
\begin{equation}
A_0 = \left(%
\begin{array}{cc}
 0  & \omega_0 \\
  - \omega_0 & 0  \\
\end{array}%
\right), \;\;\;\;
A_1 = \frac{1}{2} \left(%
\begin{array}{cc}
  -1 & i \\
  -i & -1 \\
\end{array}%
\right).
\end{equation}
and $B = (-1,i)^{T}/\sqrt{2}$.

To solve the dynamical equation (\ref{Dynamical equation of
single-mode cavity}), we use the Laplace transform
\begin{equation}\label{Laplace transform}
O \left( s \right) = \int_0^{\infty}\!\!\! \exp{ \left( - s t
\right) } O \left( t \right) d t
\end{equation}
to transform the differential equation to an algebraic equation in
the frequency domain. In this domain the solution becomes
\begin{equation}\label{Solution of linear cavity in the frequency domain}
\vec{x} \left( s \right) = \frac{ \left(%
\begin{array}{cc}
  d \left( s \right) & \omega_0 \\
  - \omega_0 & d \left( s \right) \\
\end{array}%
\right) } {\sqrt{2} \Delta \left( s
\right)} \left(%
\begin{array}{c}
  - \kappa \left( s \right) b_{\rm in} \left( s \right) - \kappa^* \left( s \right) b_{\rm in}^{\dagger} \left( s \right) \\
  i \kappa \left( s \right) b_{\rm in} \left( s \right) - i \kappa^* \left( s \right) b_{\rm in}^{\dagger} \left( s \right) \\
\end{array}%
\right),
\end{equation}
where
\begin{eqnarray*}
& \Delta \left( s \right) = d^2 \left( s \right) +
\omega_0^2,\quad d \left( s \right) = s + \gamma \left( s \right)
/ 2,
\end{eqnarray*}
and $\kappa \left( s \right)$, $\vec{x} \left( s \right)$, $b_{\rm
in} \left( s \right)$, $b_{\rm in}^{\dagger} \left( s \right) $, $
\gamma \left( s \right) $ are the Laplace transforms of $\kappa
\left( t \right)$, $\vec{x} \left( t \right) $, $b_{\rm in} \left(
t \right) $, $b_{\rm in}^{\dagger} \left( t \right)$, $\gamma
\left( t \right) $. We can also obtain the following input-output
relation in the frequency domain
\begin{equation}\label{Input-output relation of linear cavity in the frequency domain}
b_{\rm out} \left( s \right) = \frac{ s + \gamma \left( s \right)
/ 2 - \kappa^2 \left( s \right) + i \omega_0 }{ s + \gamma \left(
s \right) / 2 + i \omega_0}\; b_{\rm in} \left( s \right).
\end{equation}
Note that this result is exact as far as the frequency dependence
of the coupling to the bath is concerned. Because the system is
linear we can obtain results without deriving a master equation.
This input-output formula shows exactly how the coupling profile
applies a low-pass filter to the input field to produce the output
field.
\\[0.05cm]

\noindent {\bf Remark 3:} Let us consider a Markovian cavity with
damping rate $\gamma$, then we have $\kappa \left( s
\right)=\sqrt{\gamma}$, $\gamma \left( s \right) = \gamma$, from
which we can obtain the traditional input-output relation for a
lossy cavity from Eq.~(\ref{Input-output relation of linear cavity
in the frequency domain}) (see Eq.~(45) in Ref.~\cite{Gough2})
\begin{equation}\label{Input-output relation of a Markovian leakage cavity}
b_{\rm out} \left( s \right) = \frac{ s - \gamma/2 + i \omega_0}{
s + \gamma / 2 + i \omega_0}\; b_{\rm in} \left( s \right).
\end{equation}
\\[0.05cm]

\section{Non-Markovian quantum networks}\label{s4}

\subsection{General theory: quantum cascade systems}\label{s41}

To derive the input-output relation for complex non-Markovian
networks, we should first study the dynamics of a system composed
of two cascade-connected subsystems, also known as the ``series
product'' of two subsystems~\cite{Gough}. The Hamiltonian of two
cascaded subsystems, depicted in Fig.~\ref{Fig of the
non-Markoviann quantum cascade system} can be expressed as
\begin{eqnarray}\label{Hamiltonian of cascade systems}
H_{\rm eff} & = & H_1 + i \left\{ \left[ \int_{- \infty}^{+
\infty} \!\!\! \kappa_1 \left( \tau - t \right) b_{\rm 1,
in}^{\dagger} \left( \tau \right) d \tau \right] L_1 - {\rm
h.c.} \right\} \nonumber \\
& + & H_2 + i \left\{ \left[ \int_{- \infty}^{+ \infty} \!\!\!
\kappa_2 \left( \tau-t \right) b_{\rm 2, in}^{\dagger} \left( \tau
\right) d \tau
\right] L_2 - {\rm h.c.} \right\}, \nonumber\\
\end{eqnarray}
where $H_{i=1,2}$ and $L_{i=1,2}$ are the free Hamiltonian and
dissipation operator of the $i$-th subsystem; and $\kappa_i \left(
t \right)$ is the corresponding coupling strength between the
$i$-th subsystem and the $i$-th input field. If we omit the time
delay for the quantum field transmitting between the two
input-output components, then we have
\begin{equation}\label{Cascade input-output relation}
b_{\rm 2, in} \left( t \right) = b_{\rm 1, out} \left( t \right) =
b_{\rm 1, in} \left( t \right) + \int_0^t\!\!\! \kappa_1 \left( t
- \tau \right) L_1 \left( \tau \right) d \tau.
\end{equation}
Substituting Eq.~(\ref{Cascade input-output relation}) into
Eq.~(\ref{Hamiltonian of cascade systems}), we have
\begin{eqnarray*}
H_{\rm eff} & = & H_1 + H_2 + H_{12} + i \sum_{j=1,2}\left[ \tilde{b}_{j,\rm in}^{\dagger} L_j - L_j^{\dagger} \tilde{b}_{j, \rm in} \right]
\end{eqnarray*}
where
\begin{equation}\label{H12}
H_{12} = - i \int_0^t\!\!\! \left[ \gamma_{12}^{ \theta } \left(
\tau - t \right) L_2 L_1^{\dagger} \left( \tau \right) - {\rm
h.c.} \right] d \tau
\end{equation}
is the interaction Hamiltonian between the two subsystems
introduced by the transmitting field; the parameter
$\gamma_{12}^{\theta} \left( \tau - t \right)$ is defined by
\begin{eqnarray*}
\gamma_{12}^{\theta} \left( t - \tilde{t} \right) = \int_{ -
\infty }^{ + \infty} \kappa_1^* \left( \tau - t \right) \kappa_2
\left( \tau - \tilde{t} \right) \theta \left( t - \tau \right) d
\tau;
\end{eqnarray*}
and $\theta\! \left( t \right) $ is the step function
\begin{equation}\label{Step function}
\theta\! \left( t \right) = \left\{%
\begin{array}{ll}
    1, & t \geq 0; \\
    0, & t < 0. \\
\end{array}%
\right.
\end{equation}
The two equivalent non-Markovian input fields that interact directly with
the two subsystems via the dissipation operators $L_1$ and $L_2$
are defined as
\begin{eqnarray*}
\tilde{b}_{l, {\rm in}} \left( t \right) = \int_{- \infty}^{+
\infty}\!\!\! \kappa_l \left( t - \tau \right) b_{\rm in} \left(
\tau \right) d \tau,
\end{eqnarray*}
and these satisfy the following commutation relation
\begin{equation}\label{Commutation relation for non-Markovian input noise}
\left[ \tilde{b}_{l, {\rm in}} \left( t \right), \tilde{b}_{r,
{\rm in}}^{\dagger} \right] = \gamma_{lr} \left( t - \tilde{t}
\right),
\end{equation}
where
\begin{equation}\label{Damping rate}
\gamma_{lr} \left( t - \tilde{t} \right) = \int_{- \infty}^{+
\infty}\!\!\! \kappa_r^* \left( t - \tau \right) \kappa_l \left(
\tilde{t} - \tau \right) d \tau.
\end{equation}
\begin{figure}[t]
\centerline{\includegraphics[bb=30 383 469 509, width=8.56
cm]{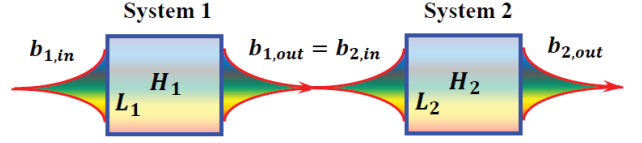}} \caption{(Color online) Schematic diagram of the
non-Markovian quantum cascade system. The output from the first
subsystem is fed into the input of the second
subsystem.}\label{Fig of the non-Markoviann quantum cascade
system}
\end{figure}
The dynamics of the total system can then be expressed as the
following quantum stochastic differential equation
\begin{eqnarray}\label{Quantum stochastic differential equation: two cascaded subsystem}
\dot{X} & = & \sum_{l,r=1}^2 \int_0^t \left\{ \gamma_{lr} \left( t
- \tau \right) L_l^{\dagger} \left( \tau \right) \left[ X \left( t
\right), L_r \left( t \right) \right] \right.
\nonumber \\
& & \left. + \gamma_{lr}^* \left( t - \tau \right) \left[
L_r^{\dagger} \left( t \right), X \left( t \right) \right] L_l
\left( \tau
\right) \right\} d \tau \nonumber \\
& & -i \left[ X, H_S \right] + \sum_{l=1}^n \left\{
\tilde{b}_{l,{\rm in}} \left[ L_l^{\dagger}, X \right] + \left[ X,
L_l \right] \tilde{b}_{l, {\rm in}}^{\dagger} \right\}, \nonumber
\\
\end{eqnarray}
where $H_S = H_1 + H_2 + H_{12}$. The input-output equation of
the cascade system can be expressed as
\begin{eqnarray}\label{Input-output relation: two cascaded subsystem}
b_{\rm out}\left( t \right) & = & b_{\rm in} \left( t \right) +
\int_0^t\!\!\! \kappa_1 \left( t - \tau \right) L_1 \left( \tau
\right) d \tau \nonumber \\
& & + \int_0^t\!\!\! \kappa_2 \left( t - \tau \right) L_2 \left(
\tau \right) d \tau.
\end{eqnarray}
\\[0.05cm]

\noindent {\bf Remark 4:} In the Markovian limit, we have
$\kappa_l \left( t \right) = \sqrt{\gamma_l} \delta \left( t
\right)$, $\gamma_{lr}^{\theta} \left( t - \tilde{t} \right) =
\gamma_{lr} \left( t - \tilde{t} \right) = \sqrt{ \gamma_l
\gamma_r } \delta \left( t - \tilde{t} \right) $, and
$\tilde{b}_{l, {\rm in}} = b_{\rm in}$. Thus, the dynamical
equation (\ref{Quantum stochastic differential equation: n
cascaded subsystem}) can be re-expressed as
\begin{eqnarray}\label{Quantum stochastic differential equation for cascade Markovian system}
\dot{X} & = & - i \left[ X, H_S \right] + \frac{1}{2} \left\{
L^{\dagger} \left[ X, L \right] + \left[ L^{\dagger}, X \right] L
\right\} \nonumber \\
& & b_{\rm in} \left[ L^{\dagger}, X \right] + \left[ X, L \right]
b_{\rm in}^{\dagger},
\end{eqnarray}
where
\begin{eqnarray*}
H_S & = & H_1 + H_2 + \frac{i \sqrt{ \gamma_1 \gamma_2 } }{2} \left( L_1^{\dagger} L_2 - L_1 L_2^{\dagger} \right) \\
L & = & \sqrt{ \gamma_1 } L_1 + \sqrt{ \gamma_2 } L_2,
\end{eqnarray*}
and output equation (\ref{Input-output relation: n cascaded
subsystem}) can be rewritten as
\begin{equation}\label{Output equation for cascade Markovian system}
b_{\rm out} \left( t \right) = b_{\rm in} \left( t \right) +
\sqrt{ \gamma_1 } L_1 \left( t \right) + \sqrt{\gamma_2} L_2
\left( t \right).
\end{equation}
These equations coincide with those obtained for the Markovian
series product systems in the literatures (see, e.g.,
Ref.~\cite{Gough}). This concludes this remark.
\\[0.05cm]

The dynamical equation (\ref{Quantum stochastic differential
equation: two cascaded subsystem}) can be extended readily to
$n$ cascade-connected subsystems to obtain the following quantum
stochastic differential equation
\begin{eqnarray}\label{Quantum stochastic differential equation: n cascaded subsystem}
\dot{X} & = & \sum_{l,r=1}^n \int_0^t\!\!\! \left\{ \gamma_{lr}
\left( t-\tau \right) L_l^{\dagger} \left( \tau \right) \left[ X
\left( t \right), L_r \left( t \right) \right] \right.
\nonumber \\
& & \left. + \gamma_{lr}^* \left( t - \tau \right) \left[
L_r^{\dagger} \left( t \right), X \left( t \right) \right] L_l
\left( \tau
\right) \right\} d \tau \nonumber \\
& & -i \left[ X, H_S \right] + \sum_{l=1}^n \left\{
\tilde{b}_{l,{\rm in}} \left[ L_l^{\dagger}, X \right] + \left[ X,
L_l \right] \tilde{b}_{l, {\rm in}}^{\dagger} \right\}, \nonumber
\\
\end{eqnarray}
where $\gamma_{lr} \left( t - \tilde{ t } \right) $ is defined by
Eq.~(\ref{Damping rate});
\begin{equation}\label{Hamiltonian of n cascade system}
H_S = \sum_{l=1}^n H_l + \sum_{l<r} H_{lr};
\end{equation}
$H_{l=1,\cdots,n}$ is the free Hamiltonian of the $l$-th
subsystem; and $H_{lr}$ is the field-mediated interaction
Hamiltonian
\begin{equation}\label{Hlr}
H_{lr} = i \int_0^t\!\!\! \left[ \gamma_{lr}^{\theta} \left( \tau
- t \right) L_r L_l^{\dagger} \left( \tau \right) - {\rm h.c.}
\right] d \tau.
\end{equation}
The function $\gamma_{lr}^{\theta} \left( t - \tilde{t} \right)$
is defined by
\begin{eqnarray*}
\gamma_{lr}^{\theta} \left( t - \tilde{ t } \right) = \int_{ -
\infty }^{ + \infty }\!\!\! \kappa_l^* \left( \tau - t \right)
\kappa_r \left( \tau - \tilde{ t } \right) \theta \left( t - \tau
\right) d \tau ,
\end{eqnarray*}
where $ \theta \left( t \right) $ is the step function defined by
Eq.~(\ref{Step function}). The output equation can be written as
\begin{eqnarray}\label{Input-output relation: n cascaded subsystem}
b_{\rm out}\left( t \right) & = & b_{\rm in} \left( t \right) +
\sum_{l=1}^n \int_0^t\!\!\! \kappa_l \left( t - \tau \right) L_l
\left( \tau \right) d \tau.
\end{eqnarray}
Transforming this into the Schr\"{o}dinger picture, we can obtain the following
second-order master equation
\begin{widetext}
\begin{eqnarray}\label{Master equation of non-Markovian series-product system}
\dot{\rho} & = & -i \left[ H_S, \rho \right] + \sum_{l,r=1}^n
\int_0^t\!\!\! \left\{ \gamma_{lr} \left( t - \tau \right)
\nonumber
 \left[ L_r \rho \left( \tau \right),
L_{ {\rm H_S}, l }^{\dagger} \left( \tau - t \right) \right] + \gamma_{lr}^* \left( t - \tau \right) \left[ L_{ {\rm
H_S}, l} \left( \tau - t \right), \rho \left( \tau \right)
L_r^{\dagger} \right] \right\} d \tau,
\end{eqnarray}
\end{widetext}
where
\begin{eqnarray*}
L_{ {\rm H_S}, l } \left( t \right) = \exp{ \left( i H_S t \right)
} L_l \exp{ \left( -i H_S t \right) }.
\end{eqnarray*}

\subsection{Example: non-Markovian qubit networks in superconducting circuits}\label{s43}

As the first example, we apply our non-Markovian network formalism
to superconducting circuits~\cite{SuperC}. Here, as a simple
example, we consider how to couple two distant single Cooper pair
boxes (CPBs) by a microwave field. As shown in Fig.~\ref{Fig of
the superconducting non-Markovian circuits}, to suppress the
decoherence effects, we embed two CPBs into two superconducting
transmission line resonators (TLRs). When we average over the
degrees of freedom of the TLRs, the interactions between the CPBs
and the input fields become non-Markovian. This can be understood
by noting that the TLRs work as microwave cavities, and these act
as low-pass filters. The white-noise input fields are filtered by
the TLRs and changed into non-Markovian Lorentz noises, and these
in turn interact with the CPBs. Thus, the qubit network considered
here is a typical non-Markovian quantum network.

The Hamiltonian of the $j^{\mbox{\textit{\scriptsize th}}}$ CPB ($j=1,2$) can be represented as
\begin{equation}\label{Hamiltonian of CPB}
H_{\rm CPB,j}=4 E_C \left( n_j -n_{gj} \right)^2 - E_J \left( \Phi_{xj} \right) \cos \phi_j,
\end{equation}
where $\phi_j$ denotes the phase drop across the
$j^{\mbox{\textit{\scriptsize th}}}$ CPB, with $n_j=-i \partial/
\left( \partial \phi_j \right)$ as its conjugate operator. The
operator $n_j$ represents the number of Cooper pairs on the island
electrode. The scalar $n_{gj}$ is the reduced charge number on the
control gate in units of Cooper pairs. This is given by
$n_{gj}=-C_g V_{gj} / 2e$, where $C_g$ and $V_{gj}$ are the gate
capacitance and gate voltage of the $j^{\mbox{\textit{\scriptsize
th}}}$ CPB. The scalar $E_C=e^2/2\left( C_g + 2 C_J^0 \right)$ is
the single-electron charging energy of a single CPB and $C_J^0$ is
the capacitance of a single Josephson junction. The Josephson
energy $E_J \left( \Phi_{xj} \right)$ of the
$j^{\mbox{\textit{\scriptsize th}}}$ DC superconducting quantum
interference device (SQUID) can be calculated by
\begin{equation}\label{Josephson energy of single Cooper pair box}
E_J \left( \Phi_{xj} \right) = 2 E_J^0 \cos \left( \pi \frac{\Phi_{xj}}{\Phi_0} \right),
\end{equation}
where $E_J^0$ represents the Josephson energy of a single
Josephson junction; $\Phi_{xj}$ is the external flux piercing the
SQUID loop of the $j^{\mbox{\textit{\scriptsize th}}}$ CPB; and
$\Phi_0$ is the flux quantum. For simplicity, we assume that $E_C$
and $E_J^0$ are the same for each Josephson junction in the two
CPBs.
\begin{figure}[t]
\includegraphics[width=8.8 cm, clip]{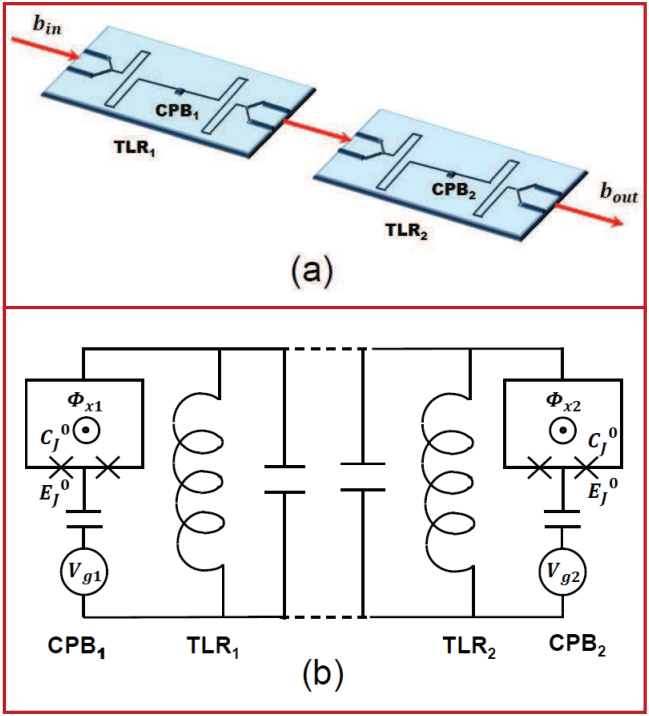}
\caption{(Color online) Schematic diagrams of (a) two
cascade-connected CPB-TLR (Cooper-pair box and transmission line
resonator) input-output systems and (b) the equivalent
superconducting circuits.}\label{Fig of the superconducting
non-Markovian circuits}
\end{figure}

Near the charge-degenerate point with $n_{gj}=0.5$, the two
lowest-energy levels of the $j^{\mbox{\textit{\scriptsize th}}}$
CPB are close to each other and far separated from higher-energy
levels. Thus, we can approximately consider a single CPB as a
two-level system. In the qubit basis, the Hamiltonian $H_{\rm
CPB,j}$ can be diagonalized as
\begin{equation}\label{Hamiltonian of the CPB in the qubit states}
H_{\rm CPB,j}=\frac{\tilde{\omega}_{qj}}{2} \sigma_z^{\left( j
\right)},
\end{equation}
where $\tilde{\omega}_{qj}=\left( E_{Cj}^2 + E_{Jj}^2
\right)^{1/2}$, $E_{Jj}=E_J \left( \Phi_{xj} \right)$ and $
E_{Cj}=4 E_C \left( 1 - 2 n_{gj} \right) $.

As shown in Fig.~\ref{Fig of the superconducting non-Markovian
circuits}, the $j^{\mbox{\textit{\scriptsize th}}}$ CPB is
capacitively coupled to the $j^{\mbox{\textit{\scriptsize th}}}$
TLR. The Hamiltonian of the $j^{\mbox{\textit{\scriptsize th}}}$
coupled CPB-TLR system can be written as
\begin{eqnarray*}
\tilde{H}_j & = & g_j \left( - \cos \alpha_j \sigma_z^{ \left( j
\right) } + \sin \alpha_j \sigma_x^{ \left( j \right) } \right)
\left( a_j + a_j^{\dagger} \right) \\
&& +\frac{\tilde{\omega}_{qj}}{2} \sigma_z^{ \left( j \right) } +
\omega_{cj} a_j^{\dagger} a_j,
\end{eqnarray*}
where $a_j$ is the annihilation operator of the single-mode
electromagnetic field in the $j^{\mbox{\textit{\scriptsize th}}}$
TLR; $g=-e \left( C_g / C_{\Sigma} \right) V_{\rm rms,j}^0$ is the
coupling strength between the resonator and the qubit;
$C_{\Sigma}=C_g+2C_J^0$ is the total capacitance of the SCB;
$V_{\rm rms}^0=\sqrt{\omega_c / 2 C_r}$ is the root mean square
(rms) of the voltage across the $LC$ circuit with $C_r$
representing the capacitance of the resonator; and $\alpha_j$ is
defined by
\begin{eqnarray*}
\alpha_j = \arctan \left[ E_J \left( \phi_{xj} \right) / E_C
\left( 1 - 2 n_{gj} \right) \right].
\end{eqnarray*}
Letting the SCBs work at the charge-degenerate point, such that
$n_{gj}=1/2$ ($j=1,2$), and introducing the rotating wave
approximation, we can obtain the following effective Hamiltonian
of the $j^{\mbox{\textit{\scriptsize th}}}$ coupled CPB-TLR system
\begin{equation}\label{Effective Hamiltonian of the CPB-TLR system}
H_j=\frac{\omega_{qj}}{2}\sigma_z^{ \left( j \right) } +
\omega_{cj} a_j^{\dagger} a_j + g_j \left( a_j^{\dagger}
\sigma_-^{ \left( j \right) } +a_j \sigma_+^{ \left( j \right) }
\right),
\end{equation}
where $\omega_{qj} = E_J \left( \phi_{xj} \right)$.

The cavity mode in the $j^{\mbox{\textit{\scriptsize th}}}$ TLR is
coupled to a transmitting field in an auxiliary transmission line
between two TLRs. The total Hamiltonian of the
$j^{\mbox{\textit{\scriptsize th}}}$ SCB-TLR system and the input
field $b_{\rm in,j }$ can be represented by
\begin{eqnarray}\label{Hamiltonian of j-th subsystem}
H_{\rm tot,j} & = & \frac{\omega_{qj}}{2} \sigma_z^{ \left( j
\right) } + \omega_{cj} a_j^{\dagger} a_j + g_j \left(
a_j^{\dagger} \sigma_-^{ \left( j \right) } + a_j \sigma_+^{
\left( j \right) } \right) \nonumber \\
&& + i \sqrt{\gamma_j} \left( a_j^{\dagger} b_{\rm in,j} - b_{\rm
in,j}^{\dagger} a_j \right).
\end{eqnarray}
where $\gamma_j$ is determined by the coupling between the cavity
mode and the input field. To be concentrated on the dynamics of
the qubit, we eliminate the degrees of freedom of the cavity mode.
From Eq.~(\ref{Hamiltonian of j-th subsystem}), it can be shown
that
\begin{equation}\label{Dynamics of transmission line resonator}
\dot{a}_j = - \left( i \omega_{cj} + \frac{\gamma_j}{2} \right)
a_j + \sqrt{\gamma_j}\; b_{\rm in,j} - i g_j \sigma_-^{ \left( j
\right) }.
\end{equation}
Let us now introduce the following weak coupling assumption
\begin{eqnarray*}
\omega_{cj},\,\gamma_j \gg g_j,
\end{eqnarray*}
so we can omit the last term in Eq.~(\ref{Dynamics of transmission
line resonator}) when we consider the dynamics of the cavity mode.
Thus, we can now solve Eq.~(\ref{Dynamics of transmission line
resonator})
\begin{equation}\label{Solution of the TLR dynamics}
a_j \left( t \right) = \sqrt{\gamma_j} \int_0^t \exp{ \left[ -
\left( i \omega_{cj}+\gamma_j/2 \right) \left( t - \tau \right)
\right] } b_{\rm j, in} \left( \tau \right) d \tau.
\end{equation}
Substituting Eq.~(\ref{Solution of the TLR dynamics}) into the
Hamiltonian $H_{\rm tot,j}$ in Eq.~(\ref{Hamiltonian of j-th
subsystem}), we can obtain an effective Hamiltonian to represent
the coupling between the $j^{\mbox{\textit{\scriptsize th}}}$
qubit and the effective input field
\begin{equation}\label{Effective Hamiltonian of j-th qubit and the input field without dring field}
\tilde{H}_{\rm eff,j} = \frac{\omega_{qj}}{2} \sigma_z^{ \left( j
\right) } + i \left( \tilde{b}_{\rm in,j}^{\dagger} \sigma_-^{
\left( j \right) } - \sigma_+^{ \left( j \right) } \tilde{b}_{\rm
in,j} \right),
\end{equation}
where
\begin{equation}\label{Effective input field interacting with the j-th
qubit} \tilde{b}_{\rm in,j} \left( t \right) = \int_0^t i g_j
\sqrt{\gamma_j} e^{ - \left( i \omega_{cj} + \gamma_j / 2 \right)
\left( t - \tau \right) } b_{\rm in,j} \left( \tau \right) d \tau.
\end{equation}
If we additionally add an ac gate voltage $V_{gj}=V_{0j} \cos
\left( \omega_{gj} t \right)$ on the gate of the
$j^{\mbox{\textit{\scriptsize th}}}$ CPB, where $V_{0j}$ and
$\omega_{gj}$ are the amplitude and frequency of the gate voltage,
we can obtain the following effective Hamiltonian in the rotating
frame
\begin{equation}\label{Effective Hamiltonian of j-th qubit and the input field}
\tilde{H}_{\rm eff,j} = \frac{\Delta_{qj}}{2} \sigma_z^{ \left( j
\right) } + i \left( \tilde{b}_{\rm in,j}^{\dagger} \sigma_-^{
\left( j \right) } - \sigma_+^{ \left( j \right) } \tilde{b}_{\rm
in,j} \right),
\end{equation}
under the condition that $C_g V_{0j} E_C / 2e \ll \Delta_{qj} =
E_J - \omega_{qj} $. By comparing Eq.~(\ref{Effective input field
interacting with the j-th qubit}) and Eq.~(\ref{Effective
Hamiltonian for non-Markovian system}), we can see that the
$j^{\mbox{\textit{\scriptsize th}}}$ CPB is just directly coupled
to the effective non-Markovian field $\tilde{b}_{\rm in,j}$ with
$\kappa_j \left( t \right) =i g_j \sqrt{ \gamma_j } \exp{ \left[ -
\left( \omega_{cj} + \gamma_j / 2 \right) t \right] } $.
Additionally we can see that the total system we consider here is
just a cascade-connected two-qubit system mediated by a
non-Markovian field. If $b_{\rm in,j}$ is a white noise, it can be
easily verified that the spectrum of $ \tilde{b}_{\rm in,j} \left(
t \right) $ is of Lorentz type. In fact, it can be calculated in
the frequency domain that
\begin{eqnarray}\label{Frequency equation of non-Markovian noise}
\tilde{b}_{\rm in,j} \left( \omega \right) & = & \kappa_j \left(
\omega \right) b_{\rm in,j} \left( \omega \right) \nonumber \\
& = & \frac{i g_j \sqrt{\gamma_j}}{\gamma_j/2 + i \left( \omega -
\omega_{cj} \right) } b_{\rm in,j} \left( \omega \right),
\end{eqnarray}
where $\tilde{b}_{\rm in,j} \left( \omega \right)$, $b_{\rm in,j}
\left( \omega \right)$, $\kappa \left( \omega \right)$ are the
Fourier transform of $b_{\rm in,j} \left( t \right) $, $b_{\rm
in,j}\left( t \right)$, $\kappa\left( t \right)$. From
Eq.~(\ref{Non-Markovian correlation rate}), it can be shown that
\begin{eqnarray*}
\tilde{\gamma}_j \left( \omega \right) = | \kappa_j \left( \omega
\right) |^2 = \frac{g_j^2 \gamma_j}{\gamma_j^2 / 4 + \left( \omega
- \omega_{cj} \right)^2}.
\end{eqnarray*}
Note that
\begin{eqnarray*}
\left[ \tilde{b}_{\rm in,j} \left( t \right), \tilde{b}_{\rm
in,j}^{\dagger} \left( \tilde{t} \right) \right] & = &
\tilde{\gamma}_j \left( t - \tilde{t} \right) = \int_0^{\infty}
e^{ - i \omega \left( t - \tilde{t} \right) } \tilde{\gamma}_j
\left( \omega \right) d
\omega \\
& = & g_j^2 \exp{ \left[ - \gamma_j \left| t - \tilde{t} \right| /
2  \right]}.
\end{eqnarray*}
We can see that $\tilde{b}_{\rm in,j} \left( t \right) $ is a
Lorentz-type noise.

To simplify our discussions, let us assume that the two qubits
have the same system parameters, i.e.,
$\Delta_q=\Delta_{q1}=\Delta_{q2}$, $g=g_1=g_2$, and
$\gamma=\gamma_1=\gamma_2$. From Eq.~(\ref{Master equation of
non-Markovian series-product system}), we can obtain the master
equation of the non-Markovian two-qubit system we consider here
\begin{eqnarray}\label{Master equation of Lorentz-type two-qubit system}
\dot{\rho} & = & -i \left[ \frac{\Delta_q}{2} J_z+ \alpha \left( t
\right) \sigma_-^{ \left( 1 \right) } \sigma_+^{ \left( 2 \right)
} + \alpha^* \left( t \right) \sigma_+^{ \left( 1 \right) }
\sigma_-^{ \left( 2 \right) }, \rho \right] \nonumber \\
&& +\int_0^t \left\{ \beta \left( t - \tau \right) \left[ J_- \rho
\left( \tau \right), J_+ \right] + {\rm h.c.} \right\} d \tau,
\end{eqnarray}
where $ J_{\alpha=z,\pm}= \sigma_{ \alpha }^{ \left( 1 \right) } +
\sigma_{ \alpha }^{ \left( 2 \right) } $ is the collective
two-qubit operator; and $\alpha \left( t \right)$ and $\beta
\left( t \right)$ can be calculated by
\begin{eqnarray}\label{alpha and beta for two-qubit example}
\alpha \left( t \right) & = & \frac{i g^2 \exp \left( i \Delta_q t
\right) }{ \gamma / 2 + i \Delta_q} - \frac{ i g^2 \exp \left( -
\gamma t / 2 \right) }{ \gamma
/ 2 + i \Delta_q }, \nonumber \\
\beta \left( t \right) & = & g^2 \exp \left[ - \gamma t /2 + i
\Delta_q t \right].
\end{eqnarray}
Additionally, if $ \Delta_q = E_J - \omega_q = 0 $, we can rewrite
Eq.~(\ref{Master equation of Lorentz-type two-qubit system}) by
introducing the first Markovian approximation, i.e., to replace
$\rho \left( \tau \right)$ in the last integral term in
Eq.~(\ref{Master equation of Lorentz-type two-qubit system}) by $
\rho \left( t \right) $, as
\begin{equation}\label{Master equation of Lorentz-type two-qubit system with the first Markovian approximation}
\dot{\rho} = -i \left[ \alpha \left( t \right) \sigma_-^{ \left( 1
\right) } \sigma_+^{ \left( 2 \right) } + {\rm h.c.}, \rho \right]
+ \Gamma \left( t \right) \mathcal{D} \left[ J_- \right] \rho,
\end{equation}
where
\begin{eqnarray*}
\alpha \left( t \right) = i \Gamma \left( t \right) = i \frac{ 2
g^2}{\gamma} \left( 1- e^{- \gamma t / 2 } \right),
\end{eqnarray*}
and the superoperator $\mathcal{D} \left[ J_- \right] \rho $ is
defined by
\begin{eqnarray*}
D \left[ J_- \right] \rho = J_- \rho J_+ -\frac{1}{2} J_+ J_- \rho
- \frac{1}{2} \rho J_+ J_-.
\end{eqnarray*}
It can be verified that both $ \alpha \left( t \right) $ and $
\Gamma \left( t \right) $ decrease when $\gamma$ decreases. This
means that both the coherent interaction between the two qubits
and the damping induced by the transmitting field decrease with
the increase of the correlation time of the non-Markovian Lorentz
noises $ \tilde{b}_{\rm in,j} \left( t \right) $ scaled by $ 1 /
\gamma $.

In Fig.~\ref{Fig of the evolution of the concurrence} we
show the evolution of the concurrence of the two qubits.
The concurrence is defined by
\begin{eqnarray*}
C \left( \rho \right) = \max \left\{ \lambda_1 - \lambda_2 -
\lambda_3 - \lambda_4, 0 \right\}.
\end{eqnarray*}
where $\rho$ is the system density matrix given by
Eq.~(\ref{Master equation of Lorentz-type two-qubit system with
the first Markovian approximation}); $\lambda_i$'s are the square
roots of the eigenvalues, in a decreasing order, of the matrix
$\rho$; and $\rho^*$ is the complex conjugate of $\rho$. From
Fig.~\ref{Fig of the evolution of the concurrence} we see that the
damping rate of the concurrence decreases as the bath coupling
strength $g$ decreases, but increases when the correlation time of
the environment ($\tau_{\rm env} = 1/\gamma$) increases. This
means that in non-Markovian environments the two-qubit
entanglement is preserved longer than in a Markovian environment,
which agrees with our intuition.

\begin{figure}[t]
\centerline{
\includegraphics[width=8.5cm]{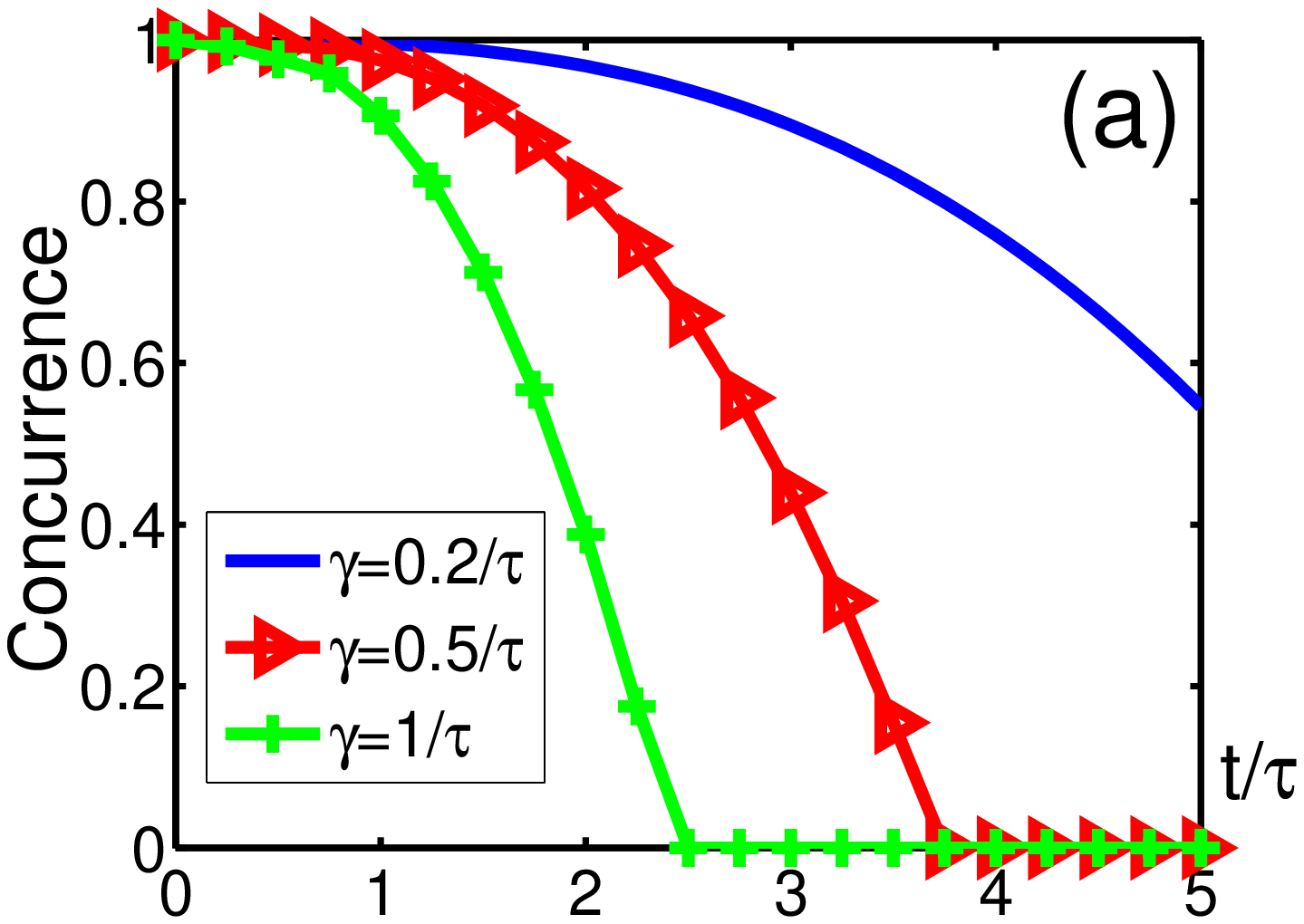}}
\centerline{
\includegraphics[width=8.5cm]{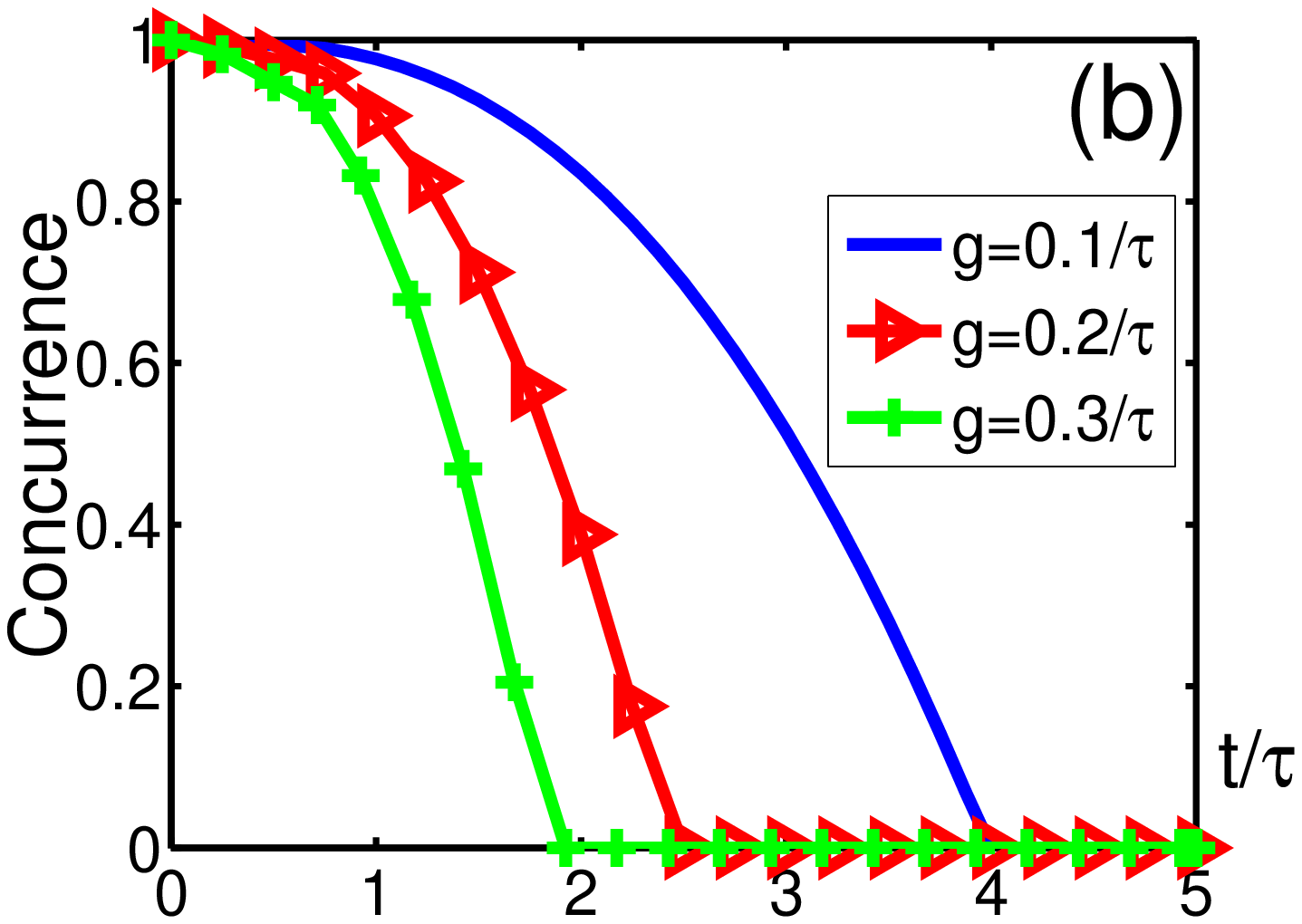}} \caption{(color online) Evolution of the
concurrence (a) for various non-Markovian correlation rates
$\gamma$ (the blue curve, red triangle curve, and the green
plus-sign curve correspond to
$\gamma=0.2/\tau,\,0.5/\tau,\,1/\tau$) and (b) for various
coupling strengths $g$ (the blue curve, red triangle curve, and
the green plus-sign curve correspond to
$g=0.1/\tau,\,0.2/\tau,\,0.3/\tau$). Here $\tau=10$ ns. The decay
of the concurrence speeds up when increasing the correlation rate
$\gamma$ of the non-Markovian noises and the qubit-environment
coupling strength $g$. }\label{Fig of the evolution of the
concurrence}
\end{figure}

The two-qubit dynamics given by Eq.~(\ref{Master equation of
Lorentz-type two-qubit system}) can be extended to
multi-qubit networks to study many-body physical phenomena, such as
quantum entanglement and correlations.

\subsection{Example: feedforward and feedback control network}\label{42}
In this subsection, we consider using networks for quantum control
~\cite{DDAlessandro, Wisemanbook, QuantumControlReview,
QuantumControl1, QuantumControl2, QFeedbackControl}. A simple
control network is composed of two cascade-connected subsystems.
One is called the \textit{controller}, and the other the
\textit{plant}. The purpose of the network is to improve the
performance of the plant by connecting it to the controller.
Whether such a strategy will work depends in general upon the
properties of each. Such a configuration is useful if for some
physical reason the controller can be tailored in ways that the
plant cannot.

It is common to divide control networks into two classes. The
first one is an open-loop or ``feedforward'' scenario in which the
input field is first fed into the controller to obtain a control
signal and then the control signal is fed into the quantum plant
to control the dynamics of the plant (see Fig.~\ref{Fig of quantum
feedforward and feedback control networks}). In this method, the
control signal contains no prior information of the system
dynamics. The second one is a feedback control
network~\cite{Jacobs2,Mabuchi2,JZhang,SLloyd,Mabuchi,Wiseman1,Doherty1,Belavkin,Jacobs1,ZHYan,YYamamoto,AYacoby,ANKorotkov,TBrandes,Jacobs3,GJMilburn,XQLi,KJPhD,JZhang2,Wiseman2}.
This is also called closed-loop control, in which the input field
is first fed into the plant to extract the information we need.
The information-bearing output field is then fed into the quantum
controller to obtain an output control signal which is fed back to
change the dynamics of the plant. Such a control system may be
viewed as a three-partite quantum cascade system:
plant-controller-plant. Since the signal fed into the controller
contains real-time information about the state of the plant, we
can use it to adjust the behavior of the controller.

By examining Eq.~(\ref{Hlr}) one can see that the
quantum controller introduces Hamiltonian terms like
\begin{equation}\label{Control Hamiltonian}
H_c = \left\{ - i \left[ \int_0^t \gamma \left( t - \tau \right)
u_c \left( \tau \right) d \tau \right] L + {\rm h.c.} \right\},
\end{equation}
where $u_c \left( t \right) = u_c \left( X_{c,1} \left( t \right),
\cdots, X_{c,n} \left( t \right) \right) $ is a function of the
system variables $ X_{c,1} \left( t \right), \cdots, X_{c,n}
\left( t \right) $ of the quantum controller and $L$ is the system
operator of the plant. The signal $u_c \left( t \right) $ is the
control signal that is input to the plant. The idea is to design
the controller dynamics so that the evolutions of $ X_{c,1} \left(
t \right), \cdots, X_{c,n} \left( t \right) $ generate the
appropriate control signal, $u_c \left( t \right)$. The primary
difference between feedforward (open-loop) and feedback
(closed-loop) controls is that in the former the controller
variables $X_{c,1}, \cdots, X_{c,n}$ can be looked as extrogenous
variables that do not depend on the dynamics of the plant. In the
latter, $X_{c,1}, \cdots, X_{c,n}$ are functions of the endogeous
variables of the plant, and can be obtained in terms of the plant
variables by solving the dynamical equation of the controller.
These two control methods have different advantages, and the best
performance might only be achieved by combing them.

\begin{figure}[t]
\includegraphics[width=8.8 cm,
clip]{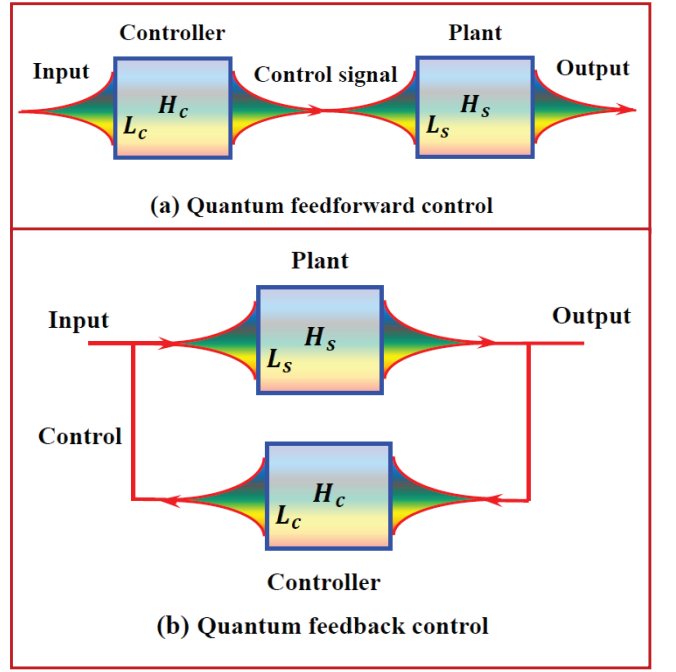}\caption{(Color online) Schematic diagrams of (a)
quantum feedforward control network and (b) quantum feedback
control network. Here, $H_c$ ($H_s$) and $L_c$ ($L_s$) are the
Hamiltonian and dissipation operator for the controller
(plant).}\label{Fig of quantum feedforward and feedback control
networks}
\end{figure}

For a concrete example, let us choose the plant to be a
single-mode cavity so that
\begin{equation}\label{Hamiltonian and dissipation operator of the plant}
H_s=\omega_a a^{\dagger} a,\quad L_s=a,
\end{equation}
and the controller to be a fully-controllable two-qubit system.
The Hamiltonian and dissipation operator of the controller are
\begin{equation}\label{Hamiltonian and dissipation operator of the controller}
H_c = \frac{\Delta_q}{ 2 } \sigma_z + \mu_d^* \sigma_- + \mu_d
\sigma_+, \quad L_c=\sigma_-.
\end{equation}
Here $\Delta_q$ is the detuning
between the transition frequency of the qubit and the frequency of
the extrogenous driving field and $\mu_d$ is a classical
extrogenous control parameter.
\begin{figure}[t]
\includegraphics[width=8.8 cm, clip]{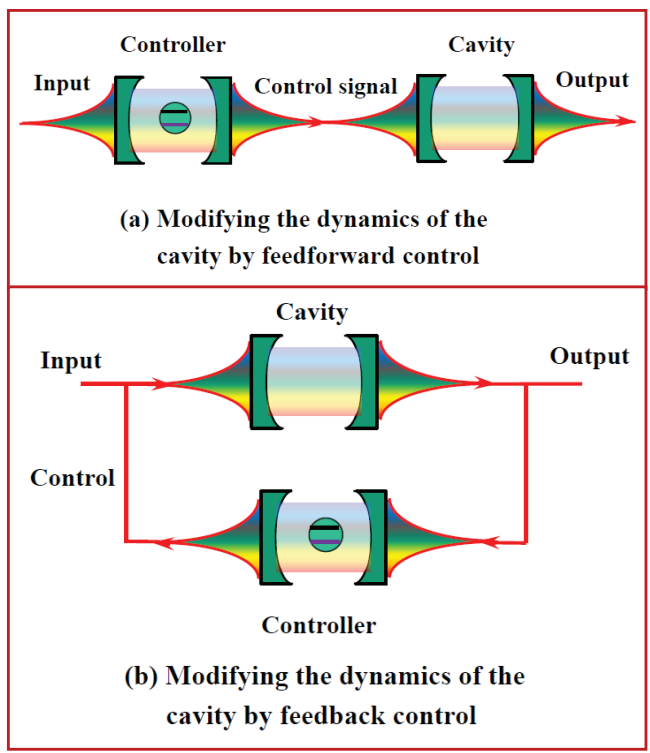}
\caption{(Color online) Modifying the dynamics of a cavity using a
qubit inside the controller by (a) feedforward control and (b)
feedback control.}\label{Fig of modifying the dynamics of a cavity
by quantum feedforward and feedback control networks}
\end{figure}

We now show what happens if we give the controller a fast damping
rate, so that it adjusts very quickly to change in the plant. In
this case we can average out the degrees of freedom of the
controller (see the derivations in Appendix~\ref{as2}). For the
case of feedforward control, the reduced Hamiltonian of the cavity
can be written as
\begin{equation}\label{Effective Hamiltonian with feedforward control}
H_{\rm 1,eff} = \omega_a a^{\dagger} a + \left[ u_c^* \left( t
\right) a + u_c \left( t \right) a^{\dagger} \right],
\end{equation}
where $u_c \left( t \right)$ is a classical control parameter
depending on the state of the qubit but does not depend on the
dynamics of the controlled cavity mode $a$. This is simply the
Hamiltonian of a driven harmonic oscillator. As a comparison, for
feedback control the reduced Hamiltonian of the cavity becomes
\begin{equation}\label{Effective Hamiltonian with feedback control}
H_{\rm 2,eff} = \omega_a a^{\dagger} a + \left\{ \left[ \int_0^t
\gamma^* \left( t - \tau \right) a^{\dagger} \left( \tau \right) d
\tau \right] a \left( t \right) + {\rm h.c.} \right\}.
\end{equation}
Non-classical optical effects such as squeezing induced by the
memory term in Eq.~(\ref{Effective Hamiltonian with feedback
control}) might well be observable in this case~\cite{Squeezing_by_coherent_feedback}.
We see that feedback control can generate a class of evolutions that are
impossible with an open-loop connection.

\section{Conclusions}\label{s5} In summary, we have extended quantum
input-output theory to arbitrary non-Markovian networks of systems
connected via continuous-wave fields. We have derived the
Heisenberg picture quantum stochastic differential equation for
the systems in the network, the corresponding perturbative master
equations, and all the input-output relations. We have applied
this general formalism to a model of two superconducting charge
qubits intracting via a cascade connection. We showed that this
system was non-Markovian because the cavities with which the
qubits connect to each other act as filters for the quantum noise.
For this system we analyzed the dynamics of the entanglement
between the qubits, and showed that it was affected by the
non-Markovian nature of the network. We also use our model to
analyze the difference between feedforward and feedback networks
in which the controller has a fast response time. It is clear from
our analysis that non-Markovian effects can have a significant
effect on the behavior of mesoscopic quantum networks, and the
analysis of these effects may be important for future quantum
devices.
\\[0.2cm]

\begin{center}
\textbf{ACKNOWLEDGMENTS}
\end{center}

J. Zhang would like to thank Prof. W.-M. Zhang for helpful
discussions. J. Zhang and R. B. Wu are supported by the National
Natural Science Foundation of China under Grant Nos. 61174084,
61134008, 60904034. Y.-X. Liu is supported by the National Natural
Science Foundation of China under Grant Nos. 10975080, 61025022.
K. Jacobs is partially supported by the NSF under Project Nos.
PHY-0902906, and  PHY-1005571, and the ARO MURI grant
W911NF-11-1-0268. F. Nori is partially supported by the ARO,
JSPS-RFBR contract No. 12-02-92100, Grant-in-Aid for Scientific
Research (S), MEXT Kakenhi on Quantum Cybernetics, and the JSPS
via its FIRST program.
\\[0.2cm]

\appendix
\section{Derivations of the non-Markovian dynamical and output
equations}\label{as1} From the system Hamiltonian (\ref{General
Hamiltonian}), we can obtain the Heisenberg equation of an
arbitrary system operator $X$
\begin{eqnarray}
\dot{X} & = & -i \left[ X, H_{\rm sys} \right] + \int d \omega
\left\{ \kappa \left( \omega \right) b^{\dagger} \left( \omega, t \right) \left[ X, L \right] \right. \nonumber \\
& & \left. - \kappa^* \left( \omega \right) \left[ X, L^{\dagger}
\right] b \left( \omega, t \right) \right\}, \label{Heisenberg
equation of the system operator}
\end{eqnarray}
and the equation of the bath operator $b \left( \omega \right)$
\begin{eqnarray}
\dot{b} \left( \omega, t \right) & = & -i \omega b \left( \omega,
t \right) + \kappa \left( \omega \right) L. \label{Heisenberg
equation of the bath operator}
\end{eqnarray}
We can solve Eq.~(\ref{Heisenberg equation of the bath operator})
and obtain
\begin{equation}\label{Bath operator in the Heisenberg picture}
b \left( \omega, t \right) = e^{- i \omega t} b \left( \omega
\right) + \kappa \left( \omega \right) \int_0^t\!\!\! e^{-i \omega
\left( t - \tau \right) } L \left( \tau \right) d \tau,
\end{equation}
where $b \left( \omega \right) = b \left( \omega, 0 \right)$ is
the initial condition of $b \left( \omega, t \right)$. Similarly,
\begin{equation}\label{Bath operator in the Heisenberg picture from future time}
b \left( \omega, t \right) = e^{- i \omega \left( t - t_1 \right)}
b \left( \omega, t_1 \right) - \kappa \left( \omega \right)
\int_t^{t_1}\!\!\! e^{-i \omega \left( t - \tau \right) } L \left(
\tau \right) d \tau,
\end{equation}
where $t_1 \geq t$. The input and output fields $b_{\rm in} \left(
t \right)$ and $b_{\rm out} \left( t \right)$ are defined as the
Fourier transform of $b \left( \omega \right)$ and $b \left(
\omega, t_1 \right)$ respectively
\begin{eqnarray*}
b_{\rm in} \left( t \right) & = & \frac{1}{\sqrt{2 \pi}} \int_{-
\infty}^{+ \infty}\!\!\! b \left( \omega \right) e^{-i \omega t}\; dt, \\
b_{\rm out} \left( t \right) & = & \frac{1}{\sqrt{2 \pi}} \int_{-
\infty}^{+ \infty}\!\!\! b \left( \omega, t_1 \right) e^{-i \omega
\left( t - t_1 \right) }\; dt.
\end{eqnarray*}
From Eqs.~(\ref{Bath operator in the Heisenberg picture}) and
(\ref{Bath operator in the Heisenberg picture from future time}),
we have
\begin{equation}\label{Output equation with t_1}
b_{\rm out} \left( t \right) = b_{\rm in} \left( t \right) +
\int_0^{t_1}\!\!\! \kappa \left( t - \tau \right) L \left( \tau
\right) d \tau.
\end{equation}
Let $t_1 \rightarrow t$, we can obtain the output
equation~(\ref{Output equation of the non-Markovian system}).

Furthermore,
using the identities
\begin{eqnarray*}
\tilde{b}_{\rm in} & = & \int_{- \infty}^{+ \infty}\!\!\! \kappa \left( t - \tau \right) b_{\rm in} \left( \tau \right) d \tau \\
& = & \int_{- \infty}^{+ \infty}\!\!\! \kappa \left( \omega
\right) e^{-i \omega t} b \left( \omega \right) d \omega,
\end{eqnarray*}
and
\begin{eqnarray*}
\gamma \left( t - \tilde{t} \right) & = & \int_{- \infty}^{+
\infty}\!\!\! \kappa^*\! \left( t - \tau \right)\; \kappa\! \left( \tilde{t} - \tau \right) d \tau \\
& = & \int_{- \infty}^{+ \infty}\!\!\! \kappa \left( \omega
\right)\; \kappa^* \left( \omega \right) e^{-i \omega \left(
\tilde{t} - t \right)}\; d \omega,
\end{eqnarray*}
we can obtain Eq.~(\ref{Quantum stochastic differential equation
for non-Markovian system}) by substituting Eq.~(\ref{Bath operator
in the Heisenberg picture}) into Eq.~(\ref{Heisenberg equation of
the system operator}).

To derive the master equation (\ref{Non-Markovian master
equation}), we first change into the interaction picture, in which
the effective Hamiltonian $H_{\rm eff}$ can be rewritten as
\begin{equation}\label{HIeff}
H_{\rm I, eff} = i \left[ b_{\rm in}^{\dagger} \left( t \right)
L_{\rm H_S} \left( t \right) - L_{\rm H_S} \left( t \right) b_{\rm
in} \left( t \right) \right],
\end{equation}
where $L_{\rm H_S} \left( t \right) $ is given in Eq.~(\ref{LHS}).
The density operator $\rho_{\rm I, tot}$ satisfies the following
Liouville equation
\begin{equation}\label{Liouville equation}
\dot{\rho}_{\rm I, tot} = - i \left[ H_{\rm I, eff} \left( t
\right), \rho_{\rm I, tot} \right].
\end{equation}
Integrating the two sides of Eq.~(\ref{Liouville equation}), we
have
\begin{equation}\label{Integrating the Liouville equation}
\rho_{\rm I, tot}\left( t \right) = -i \int_0^t \left[ H_{\rm I,
eff} \left( \tau \right), \rho_{\rm I, eff} \left( \tau \right)
\right] d \tau.
\end{equation}
Substituting Eq.~(\ref{Integrating the Liouville equation}) into
Eq.~(\ref{Liouville equation}), we can obtain
\begin{equation}\label{Interative Liouville equation}
\dot{\rho}_{\rm I, tot} = \int_0^t \left[ H_{\rm I, eff} \left( t
\right), \left[ H_{\rm I, eff} \left( \tau \right), \rho_{\rm I,
tot} \left( \tau \right) \right] \right] d \tau.
\end{equation}
Tracing over the degrees of freedom of the input field, we can
obtain the dynamical equation of the system density operator
$\rho_I = {\rm tr}_B \rho_{\rm I, tot}$
\begin{equation}\label{Master equation without Born approximation}
\dot{\rho}_I = \int_0^t {\rm tr}_B \left\{ \left[ H_{\rm I, eff}
\left( t \right), \left[ H_{\rm I, eff} \left( \tau \right),
\rho_{\rm I, tot} \left( \tau \right) \right] \right] \right\} d
\tau.
\end{equation}
Let us then introduce the Born approximation and assume that the
input field stays in the vacuum state, we have
\begin{equation}\label{Born approximation}
\rho_{\rm I, tot} \left( t \right) = \rho_I \left( t \right)
\otimes | 0 \rangle_{B\,B} \langle 0 |.
\end{equation}
Notice that it can be shown that
\begin{eqnarray}\label{Expection of the input noise}
& \langle b_{\rm in} \left( t \right) b_{\rm in}^{\dagger} \left(
\tilde{t} \right) \rangle = \delta \left( t - \tilde{t} \right), &
\nonumber \\
& \langle b_{\rm in}^{\dagger} \left( t \right) b_{\rm in} \left(
\tilde{t} \right) \rangle = \langle b_{\rm in}^{\dagger} \left( t
\right) b_{\rm in}^{\dagger} \left( \tilde{t} \right) \rangle =
\langle b_{\rm in} \left( t \right) b_{\rm in} \left( \tilde{t}
\right) \rangle = 0, & \nonumber \\
\end{eqnarray}
where $\langle \cdot \rangle $ is defined by $\langle R \rangle =
\langle 0 | b | 0 \rangle_B $. Substituting Eqs.~(\ref{HIeff}),
(\ref{Born approximation}), and (\ref{Expection of the input
noise}) into Eq.~(\ref{Master equation without Born
approximation}), we can verify that
\begin{equation}\label{Master equation with Born approximation}
\dot{\rho}_I=\int_0^t \left\{ \gamma \left( t - \tau \right)
\left[ L_{H_S} \left( t \right) \rho_I \left( \tau \right),
L_{H_S}^{\dagger} \left( \tau \right) \right] + {\rm h.c. }
\right\}.
\end{equation}
We can derive Eq.~(\ref{Non-Markovian master equation}) by
transforming Eq.~(\ref{Master equation with Born approximation})
back into the Schr\"{o}dinger picture.

\section{Derivations of the effective Hamiltonians for feedforward and feedback
control}\label{as2}

For the case of feedforward control, the dynamics of the
feedforward control system shown in Fig.~\ref{Fig of modifying the
dynamics of a cavity by quantum feedforward and feedback control
networks}(a) can be represented by
\begin{eqnarray}\label{Dynamical euqations of feedforward control system}
\dot{\sigma}_- & = & -i \Delta_q \sigma_- + i \left( \mu_d +
g_{qb} b
\right) \sigma_z, \nonumber \\
\dot{\sigma}_z & = & 2 i \left( \mu_d + g_{qb} b \right) \sigma_-
- 2 i \left( \mu_d + g_{qb} b \right) \sigma_+, \nonumber \\
\dot{b} & = & -\frac{\gamma_b}{2} b - i \omega_b b + i g_{qb}
\sigma_- + \sqrt{\gamma_b} b_{\rm in}, \nonumber \\
b_{\rm out} & = & b_{\rm in} + \sqrt{\gamma_b} b, \nonumber \\
\dot{a} & = & - \left( \frac{\gamma_a}{2} +i \omega_a \right) a +
\sqrt{\gamma_a} b_{\rm out},
\end{eqnarray}
where $b$ is the annihilation operator of the cavity mode directly
interacting with the qubit; $\gamma_a$, $\gamma_b$ are the decay
rates of the cavity modes $a,\,b$; and $g_{qb}$ is the coupling
strength between the qubit and the cavity mode $b$. Here we omit
the decay of the qubit. By averaging out the input noise, we have
\begin{equation}\label{Solution of the cavity mode b}
b = i g_{qb} \int_0^t \exp\left[ - \left( i \omega_b + \gamma_b /
2 \right) \left( t - \tau \right) \right] \sigma_- \left( \tau
\right) d \tau.
\end{equation}
Under the condition that $\mu_d \gg g_{qb} \langle b \rangle$
where $ \langle b \rangle $ is the average of the cavity mode $b$,
we can rewrite Eq.~(\ref{Dynamical euqations of feedforward
control system}) by substituting Eq.~(\ref{Solution of the cavity
mode b}) into Eq.~(\ref{Dynamical euqations of feedforward control
system})
\begin{eqnarray}\label{non-Markovian dynamical euqations of feedforward control system}
\dot{\sigma}_- & = & -i \Delta_q \sigma_- + i \mu_d \sigma_z, \nonumber \\
\dot{\sigma}_z & = & 2 i \mu_d \sigma_- - 2 i \mu_d \sigma_+, \nonumber \\
b_{\rm out} & = & b_{\rm in} + i \sqrt{\gamma_b} g_{qb} \int_0^t
e^{- \left( i \omega_b + \gamma_b / 2 \right) \left( t - \tau
\right) } \sigma_- \left( \tau \right) d \tau, \nonumber \\
\dot{a} & = & i \sqrt{ \gamma_a \gamma_b } g_{qb} \int_0^t e^{-
\left( i \omega_b + \gamma_b / 2 \right) \left( t - \tau \right) }
\sigma_- \left( \tau
\right) d \tau \nonumber \\
& & -\left( \gamma_a / 2 + i \omega_a \right) a.
\end{eqnarray}
Thus the dynamics of the cavity mode $a$ is dominated by the
effective Hamiltonian $H_{\rm 1,eff}$ given in Eq.~(\ref{Effective
Hamiltonian with feedforward control}), where
\begin{eqnarray*}
u_c = - \sqrt{\gamma_a \gamma_b} g_{qb} \int_0^t \exp\left[ -
\left( i \omega_b + \gamma_b / 2 \right) \left( t - \tau \right)
\right] \sigma_- \left( \tau \right) d \tau.
\end{eqnarray*}
Under the semiclassical approximation, $\sigma_- \left( t \right)$
can be replaced by its average. With this simplification, $u_c
\left( t \right) $ can be seen as a classical extrogenous control
parameter because the dynamics of the qubit determined by
$\sigma_-$ and $\sigma_z$ does not depend on the cavity mode $a$.
With similar discussions, we can also obtain the effective
Hamiltonian $H_{\rm 2,eff}$ induced by feedback control given in
Eq.~(\ref{Effective Hamiltonian with feedback control}).

\end{document}